\documentclass[11pt]{article}

\usepackage{amssymb}
\usepackage{amsfonts}
\usepackage{amsmath}
\usepackage[english,german]{babel}
\usepackage{epsfig}
\usepackage{latexsym}
\usepackage{color}
\usepackage{epsfig}
\usepackage{psfrag,subfigure}

\newtheorem{thm}{Theorem}[section]
\newtheorem{prop}[thm]{Proposition}
\newtheorem{cor}[thm]{Corollary}
\newtheorem{lemma}[thm]{Lemma}
\newtheorem{definition}[thm]{Definition}
\newtheorem{rem}[thm]{Remark}
\newtheorem{example}[thm]{Example}

\newenvironment{Proof}{\textsc{Proof.}}{\mbox{ } \hfill $\Box$ \vspace{2mm}}

\newenvironment{rmenumerate}
    {\begin{enumerate}}
    {\end{enumerate}}

\newcommand{\ba}{\begin{array}{ll}}
\newcommand{\bal}{\begin{array}{ll}}
\newcommand{\ea}{\end{array}}

\newcommand{\n}{\mathbb{N}}
\newcommand{\re}{\mathbb{R}}
\newcommand{\E}{\mathbb{E}}
\newcommand{\Prob}{\mathbb{P}}

\def\n{\mbox{$\mathbb{N}$}}

\def\X{\mbox{$\cal{X}$}}
\def\t{\mbox{${\theta}$}}
\def\T{\mbox{${\Theta}$}}
\def\O{\mbox{${\Omega}$}}

\newcommand{\mf}{\mathcal{F}}
\newcommand{\ml}{\mathcal{L}}

\begin{document}

\selectlanguage{english}

\renewcommand{\baselinestretch}{1.25} \normalsize


\title{Risk Minimization and Optimal Derivative Design in a Principal Agent Game\thanks{We thank Guillaume Carlier, Pierre-Andre Chiappori, Ivar Ekeland, Andreas Putz and seminar participants at various institutions for valuable comments and suggestions. Financial support through an NSERC individual discovery grant is gratefully acknowledged.}}

\author{\normalsize Ulrich Horst \\[8pt]
        \small  Department of Mathematics  \\
        \small  Humboldt University Berlin \\
        \small  Unter den Linden 6\\
        \small  10099 Berlin \\
        \small  horst@math.hu-berlin.de
         \and
        \normalsize  Santiago Moreno-Bromberg \\[8pt]
        \small Department of Mathematics\\
        \small University of British Columbia \\
        \small 1984 Mathematics Road\\
        \small Vancouver, BC, V6T 1Z2\\
        \small smoreno@math.ubc.ca
\vspace*{0.8cm}}

\maketitle

\begin{abstract}
We consider the problem of Adverse Selection and optimal
derivative design within a Principal-Agent framework. The
principal's income is exposed to non-hedgeable risk factors
arising, for instance, from weather or climate phenomena. She
evaluates her risk using a  coherent and law invariant risk
measure and tries minimize her exposure by selling derivative
securities on her income to individual agents. The agents have
mean-variance preferences with heterogeneous risk aversion
coefficients. An agent's degree of risk aversion is private
information and hidden to the principal who only knows the overall
distribution. We show that the principal's risk minimization
problem has a solution and illustrate the effects of risk transfer
on her income by means of two specific examples. Our model extends
earlier work of Barrieu and El Karoui (2005) and Carlier, Ekeland
and Touzi (2007).
\end{abstract}

\vspace{1mm}

\begin{center}
\textsc{Preliminary - Comments Welcome}
\end{center}

\vspace{1mm}


\vspace{2mm}


\hfill

\newpage

\renewcommand{\baselinestretch}{1.20} \normalsize

\section{Introduction}

In recent years there has been an increasing interest in
derivative securities at the interface of finance and insurance.
\textsl{Structured products} such as risk bonds, asset-backed
securities and weather derivatives are end-products of a process
known as \textsl{securitization} that transforms non-tradable risk
factors into tradable financial assets. Developed in the U.S.
mortgage markets, the idea of pooling and underwriting risk that
cannot be hedged through investments in the capital markets alone
has long become a key factor driving the convergence of insurance
and financial markets.

Structured products are often written on non-tradable underlyings,
tailored to the issuers specific needs and traded ``over the
counter''. Insurance companies, for instance,
routinely sell weather derivatives or risk bonds to customers that
cannot go to the capital markets directly and/or
seek financial securities with low correlation with stock indices
as additions to diversified portfolios. The market for such claims is generally
incomplete and illiquid. As a result, many of the standard paradigms of
traditional derivative pricing theory, including replication
arguments do not apply to structured products. In an illiquid
market framework, preference-based valuation principles that take
into account characteristics and endowment of trading partners may
be more appropriate for designing, pricing and hedging contingent
claims. Such valuation principles have become a major topic of
current research in economics and financial mathematics. They
include rules of Pareto optimal risk allocation (\cite{kn:fk},
\cite{kn:HIM}), market completion and dynamic equilibrium pricing
(\cite{kn:HM}, \cite{kn:HPR}) and, in particular, utility
indifference arguments (\cite{kn:BELK}, \cite{kn:BELK2},
\cite{kn:Becherer}, \cite{kn:Becherer2}, \cite{kn:Davis}, ...).
The latter assumes a high degree of market asymmetry.
For indifference valuation to be a \textsl{pricing} rather than \textsl{valuation}
principle, the demand for a financial security must come from identical agents
with known preferences and negligible market impact while the supply must come from a single principal.
When the demand comes from heterogeneous individuals with hidden characteristics,
indifference arguments do not always yield an appropriate pricing scheme.

In this paper we move away from the assumption
of investor homogeneity and allow for heterogeneous agents.
We consider a single principal with a random endowment whose goal
is to lay off some of her risk with heterogeneous agents by
designing and selling derivative securities on her income.  The
agents have mean variance preferences. An agent's degree of risk
aversion is private information and hidden to the principal. The
principal only knows the distribution of risk aversion
coefficients which puts her at an informational disadvantage. If
all the agents were homogeneous, the principal, when offering a
structured product to a single agent, could (perhaps) extract the
indifference (maximum) price from each trading partner. In the presence of
agent heterogeneity this is no longer possible, either because
the agents would hide their characteristics from the principal or
prefer another asset offered by the principal but designed and
priced for another
customer. 

The problem of optimal derivative design in a Principal-Agent
framework with informed agents and an uninformed principal has
first been addressed in a recent paper by of Carlier, Ekeland and
Touzi \cite{kn:cet}. With the agents being the informed party,
theirs is a screening model. The literature on screening within
the Adverse Selection framework can be traced back to Mussa and
Rosen \cite{kn:mr}, where both the principal's allocation rule and
the agents' types are one-dimensional. Armstrong \cite{kn:A}
relaxes the hypothesis of agents being characterized by a single
parameter. He shows that, unlike the one-dimensional case,
``bunching" of the first type is robust when the types of the
agents are multi-dimensional.  In their seminal paper, Rochet and
Chon\'e \cite{kn:rch} further extend this analysis. They provide a
characterization of the contracts, determined by the (non-linear)
pricing schedule, that maximize the principal's utility under the
constraints imposed by the asymmetry of information in the models.
Building on their work, Carlier, Ekeland and Touzi \cite{kn:cet}
study a Principal-Agent model of optimal derivative design where
the agents' preferences are of mean-variance type and their
multi-dimensional types characterize their risk aversions and
initial endowments. They assume that there is a direct cost to the
principal when she designs a contract for an agent, and that the
principal's aim is to maximize profits.

We start from a similar set-up, but substitute the idea that
providing products carries a cost for the idea that traded
contracts expose the principal to additional risk - as
measured by a convex risk measure - in exchange for a known
revenue. This may be viewed as a partial extension of
the work by Barrieu and El Karoui (\cite{kn:BELK},\cite{kn:BELK2})
to an incomplete information framework.

The principal's aim is to
minimize her risk exposure by trading with the agents subject to
the standard incentive compatibility and individual rationality
conditions on the agents' choices. In order to prove that the
principal's risk minimization problem has a solution we first
follow the seminal idea of Rochet and Chon\'e \cite{kn:rch} and
characterize incentive compatible catalogues in terms of
$U$-convex functions. When the impact of a single trade on the
principal's revenues is linear as in Carlier, Ekeland and Touzi
\cite{kn:cet}, the link between incentive compatibility and
$U$-convexity is key to establish the existence of an optimal
solution. In our model the impact is non-linear as a single trade
has a non-linear impact on the principal's risk assessment. Due to
this non-linearity we face a non-standard variational problem
where the objective cannot be written as the integral of a given
Lagrangian. Instead, our problem can be decomposed into a standard
variational part representing the aggregate income of the
principal, plus the minimization of the principal's risk
evaluation, which depends on the aggregate of the derivatives
traded. We state sufficient conditions that guarantee that the principal's optimization problem has a solution and illustrate the effect of risk transfer on her exposure by means of two specific examples.

The remainder of this paper is organized as follows. In Section
\ref{Setup} we formulate our Principal-Agent model and state the
main result. The proof is given in Section \ref{Main}. In Section
\ref{sec-examples} we illustrate the effects of risk transfer on
the principal's position by two examples. In the first we
consider a situation where the principal restricts itself to
type-dependent multiples of some benchmark claim. This case can be
solved in closed form by means of a standard variational problem.
The second example considers put options with type-dependent
strikes. In both cases we assume that the principal's risk measure
is Average Value at Risk. As a consequence the risk minimization
problem can be stated in terms of a min-max problem; we provide
an efficient numerical scheme for approximating the optimal
solution. The code is given in an appendix.


\section{The Microeconomic Setup}\label{Setup}

We consider an economy with a single {\sl principal} whose income
$W$ is exposed to non-hedgeable risk factors rising from, e.g.,
climate or weather phenomena. The random variable $W$ is defined
on a standard, non-atomic, probability space
$\left(\Omega,{\cal{F}}, \Prob \right)$ and it is square
integrable:
\[
    W \in L^{2}(\Omega,{\cal{F}}, \Prob).
\]

The principal's goal is to lay off parts of her risk with
individual {\sl agents}. The agents have heterogenous
mean-variance preferences\footnote{Our analysis carries over to preferences of mean-variance type with random initial endowment as in \cite{kn:cet}; the assumption of simple mean-variance preferences is made for notational convenience.} and are indexed by their coefficients of
risk aversion  $\theta \in \Theta$. Given a contingent claim $Y
\in L^{2}(\Omega,{\cal{F}}, \Prob)$ an agent of type $\theta$
enjoys the utility
\begin{equation}
    U(\t, Y)=\E[Y]-\t\, \textnormal{Var} [Y].
\end{equation}
Types are private information. The principal knows the
distribution $\mu$ of types but not the realizations
of the random variables $\theta$. We assume that the agents are
risk averse and that the risk aversion coefficients are bounded
away from zero. More precisely,
\[
    \Theta = [a,1] \qquad \mbox{for some } a > 0.
\]

The principal offers a derivative security $X(\t)$ written on her random
income for any type $\t$. The set of all such securities is denoted by
\begin{equation}\label{eq:1}
    \X:=\left\{X = \{X(\t)\}_{\t \in \T} \,\mid\, X \in L^{2}(\O\times\T, \Prob \otimes\mu), \,
    X\,\,{\mbox{is}}\,\,\sigma(W)\times{\cal{B}}(\T)\,\,{\mbox{measurable}}\right\}.
\end{equation}
We refer to a list of securities $\{X(\t)\}$ as a {\it contract}.
A {\it catalogue} is a contract along with prices $\pi(\t)$ for
every available derivative $X(\t)$. For a given catalogue $(X,\pi)$
the optimal net utility of the agent of type $\t$ is given by
\begin{equation} \label{v-def}
    v(\t) = \sup_{\t ' \in \T} \left\{ U(\t,X(\t ')) - \pi(\t') \right\}.
\end{equation}

\begin{rem}
No assumption will be made on the sign of $\pi(\t);$ our model
contemplates both the case where the principal takes additional
risk in exchange of financial compensation and the one where she
pays the agents to take part of her risk.
\end{rem}

A catalogue $(X,\pi)$ will be called {\it incentive compatible}
({\bf{IC}}) if the agent's interests are best served by revealing
her type. This means that her optimal utility is achieved by the
security $X(\theta)$:
\begin{equation}
    U(\t, X(\t))-\pi(\t)\ge U(\t, X(\t'))-\pi(\t') \quad \mbox{for all} \quad \t,\,\t' \in\T.
\end{equation}

We assume that each agent has some outside option (``no trade'')
that yields a utility of zero. A catalogue is thus called {\it
individually rational} ({\bf{IR}}) if it yields at least the
reservation utility for all agents, i.e., if
\begin{equation}
    U(\t, X(\t))-\pi(\t)\ge 0 \quad \mbox{for all} \quad \t \in \T.
\end{equation}

\begin{rem} By offering only incentive compatible contracts, the
principal forces the agents to reveal their type. Offering
contracts where the IR constraint is binding allows the principal
to exclude undesirable agents from participating in the market. It
can be shown that under certain conditions, the interests of the
Principal are better served by keeping agents of ``lower types" to
their reservation utility; 
Rochet and Chon\'{e} \cite{kn:rch}
have shown that in higher dimensions this
is always the case.
\end{rem}

If the principal issues the catalogue $(X, \pi)$, she receives a
cash amount of $\int_{\T} \pi \left(\theta \right)d \mu ( \theta)$
and is subject to the additional liability $\int_{\T} X(\theta)
\mu (d\t).$ She evaluates the risk associated with her overall
position
\[
    W + \int_{\T}(\pi(\t)-X(\t))d\mu(\t)
\]
via a coherent and law-invariant risk measure $\varrho:
L^2(\Omega, \mf, \Prob) \rightarrow \re \cup \{\infty\}$ that has
the Fatou property. It turns out that such risk measures can be represented as robust mixtures of Average Value at Risk.\footnote{We review properties of coherent risk measures on $L^p$ spaces in the appendix and refer to the textbook by F\"ollmer and Schied \cite{kn:fs} and the paper of Jouini, Schachermayer and Touzi \cite{kn:jst} for detailed discussion of law invariant risk measures.}
The principal's risk associated
with the catalogue $(X, \pi)$ is given by
\begin{equation}\label{eq:R}
    \varrho\left(W + \int_{\T}(\pi(\t)-X(\t)) d\mu(\t)\right).
\end{equation}
Her goal is to devise contracts $(X,\pi)$ that minimize
(\ref{eq:R}) subject to the incentive compatibility and individual
rationality condition:
\begin{equation} \label{Principal-Problem}
    \inf \left\{
    \varrho\left(W + \int_{\T}(\pi(\t)-X(\t)) d\mu(\t)\right) \, \mid \,
    X \in \X, \, X \mbox{ is {\bf IC} and {\bf IR}} \right\}.
\end{equation}

We are now ready to state the main result of this paper. The proof
requires some preparation and will be carried out in the following
section.

\begin{thm} \label{Main-thm}
If $\varrho$ is a coherent and law invariant  risk measure on
$L^2(\Prob)$ and if $\varrho$ has the Fatou property, then the
principal's optimization problem has a solution.
\end{thm}

For notational convenient we establish our main result for the spacial case
$d\mu(\t)=d\t$. The general case follows from straight forward modifications.


\section{Proof of the Main Theorem}\label{Main}

Let $(X,\pi)$ be a catalogue. In order to prove our main result it will be convenient to assume
that the principal offers any square
integrable contingent claim and to view the agents' optimization
problem as optimization problems over the set $L^2(\Prob)$. This
can be achieved by identifying the price list $\{\pi(\t)\}$ with
the pricing scheme
\[
    \pi: L^2(\Prob) \rightarrow \re
\]
that assigns the value $\pi(\t)$ to an available claim $X(\t)$ and
the value $\E[Y]$ to any other claim $Y \in
L^2$. In terms of this pricing scheme the value function $v$
defined in (\ref{v-def}) satisfies
\begin{equation} \label{v-def2}
    v(\t) = \sup_{Y \in L^2(\Prob)} \left\{ U(\t,Y) - \pi(Y) \right\}.
\end{equation}
for any individually rational catalogue. For the remainder of this section we shall work with the value
function of the (\ref{v-def2}). It is $U$-convex in the sense of
the following definition; it actually turns out to be convex and
non-increasing as we shall prove in Proposition \ref{pr:Uconv}
below.

\begin{definition}
Let two spaces $A$ and $B$ and a function $U :A\times B\to\re$ be given.
\begin{rmenumerate}
\item The function $f:A\to\re$ is called $U$-convex if there exists a function $p: B \to \re$ such that 
\[
    f(a)=\sup_{b\in B}\left\{U(a,b)-p(b)\right\}.
\]
\item For a given function $p:B \to \re$ the $U$-conjugate $p^U(a)$ of $p$ is defined by
\[
    p^U(a) = \sup_{b\in B}\left\{U(a,b)-p(b)\right\}.
\]
\item The {\it {U-subdifferential}} of $p$ at $b$ is given by the set
\[
    \partial_U p(b):=\left\{a\in A\,\mid\, p^U(a)= U(a,b)-p(b)\right\}.
\]
\item
If $a\in \partial_U p(b),$ then $a$ is called a {\it {U-subgradient}} of $p(b).$
\end{rmenumerate}
\end{definition}

Our goal is to identify the class of {\bf IC} and {\bf IR}
catalogues with a class of convex and non-increasing functions on
the type space. To this end, we first recall the link between
incentive compatible contracts and $U$-convex functions from Rochet and Chon\'e \cite{kn:rch} and Carlier, Ekeland and Touzi \cite{kn:cet}.

\begin{prop} (\cite{kn:rch}, \cite{kn:cet}) \label{pr:Uconv} If  a catalogue $(X, \pi)$ is incentive compatible, then the
function $v$ defined by (\ref{v-def}) is proper  and U-convex and
$X(\t)\in\partial_Uv(\t).$ Conversely,  any proper, U-convex
function induces an incentive compatible catalogue.
\end{prop}
\begin{Proof} Incentive compatibility of a catalogue $(X, \pi)$ means that
\[
    U(\t, X(\t))-\pi(\t)\ge U(\t, X(\t'))-\pi(\t') \quad \mbox{for all} \quad \t,\,\t' \in\T,
\]
so $v(\t)=U(\t, X(\t))-\pi(\t)$ is U-convex and
$X(\t)\in\partial_Uv(\t).$ Conversely, for a proper, U-convex
function $v$ and $X(\t)\in\partial_Uv(\t)$ let
\[
    \pi(\t):=U(\t, X(\t))-v(\t).
\]
By the definition of the U-subdifferential, the catalogue $(X,
\pi)$ is incentive compatible.
\end{Proof}

The following lemma is key. It shows that the $U$-convex function
$v$ is convex and non-increasing and that any convex and
non-increasing function is $U$-convex, i.e., it allows a
representation of the form (\ref{v-def2}). This allows us to
rephrase the principal's problem as an optimization problem over a
compact set of convex functions.

\begin{lemma}\label{lm:1}
\begin{rmenumerate}
\item Suppose that the value function $v$ as defined by
(\ref{v-def2}) is proper. Then $v$ is convex and non-increasing.
Any optimal claim $X^*(\t)$ is a $U$-subgradient of
$v(\t)$ and almost surely
    \[
        -\textnormal{Var}[X^*(\t)] = v'(\t).
    \]
\item If $\bar{v}: \T \to \re_+$ is proper, convex and
non-increasing, then $\bar{v}$ is $U$-convex, i.e., there exists a
map $\bar{\pi}: L^2(\Prob) \to \re$ such that
    \[
        \bar{v}(\t) = \sup_{Y \in L^2(\Prob)} \left\{ U(\t,Y) - \bar{\pi}(Y) \right\}.
    \]
    Furthermore, any optimal claim $\bar{X}(\t)$ belongs to the $U$-subdifferential of $\bar{v}(\t)$ and satisfies
    \[
        -\textnormal{Var}[\bar{X}(\t)] = \bar{v}'(\t).
    \]
\end{rmenumerate}
\end{lemma}
\begin{Proof}
\begin{rmenumerate}
\item Let $v$ be a  proper, $U$-convex function. Its $U$-conjugate is:
\begin{eqnarray*}
    v^U(Y) &=& \sup_{\t\in\T} \left\{\E[Y]-\t \textnormal{Var}[Y]-v(\t)\right\}\\
    &=& \E[Y]+\sup_{\t\in\T} \left\{\t(-\textnormal{Var}[Y])-v(\t)\right\}\\
    &=& \E[Y] + v^*(-\textnormal{Var}[Y]),
\end{eqnarray*}
where $v^*$ denotes the convex conjugate of $v.$ As a $U$-convex
function, the map $v$ is characterized by the fact that
$v=(v^U)^U$. Thus
\begin{eqnarray*}
    v(\t) &=& (v^U)^U(\t) \\
    &=& \sup_{Y \in L^2(\Prob)}\left\{U(\t,Y)-\E[Y]-v^*(-\textnormal{Var}[Y])\right\}\\
    &=& \sup_{Y \in L^2(\Prob)}\left\{\E[Y]-\t \textnormal{Var}[Y]-\E[Y]-v^*(-\textnormal{Var}[Y])\right\}\\
    &=& \sup_{y\le 0}\left\{\t\cdot y-v^*(y)\right\}
\end{eqnarray*}
where the last equality uses the fact that the agents' consumption
set contains claims of any variance. We deduce from the preceding representation that $v$ is
non-increasing. Furthermore $v = (v^*)^*$ so $v$ is convex. To
characterize $\partial_U v(\t)$ we proceed as follows:
\begin{eqnarray*}
    \partial_U v(\t) &=& \left\{Y \in L^2 \,\mid\, v(\t)=U(\t, X)-v^U(Y)\right\}\\
    &=&\left\{Y \in L^2 \,\mid\, v(\t)= \E[Y]-\t \textnormal{Var}[Y]-v^U(Y)\right\}\\
    &=&\left\{Y \in L^2 \,\mid\, v(\t)= \E[Y]-\t \textnormal{Var}[Y]-\E[Y]-v^*(-\textnormal{Var}[Y])\right\}\\
    &=&\left\{Y \in L^2 \,\mid\, v(\t)=\t (-\textnormal{Var}[Y])-v^*(-\textnormal{Var}[Y])\right\}\\
    &=&\left\{Y \in L^2 \,\mid\, -\textnormal{Var}[Y]\in\partial v(\t)\right\}
\end{eqnarray*}
The convexity of $v$ implies it is a.e. differentiable so we may
write
\[
    \partial_U v(\t):=\left\{Y \in L^2\,\mid\, v'(\t)=-\textnormal{Var}[Y])\right\}.
\]

\item Let us now consider a proper, non-negative, convex and
non-increasing function $\bar{v}: \T \to \re$. There exists a map
$f:\re\to\re$ such that
\[
    \bar{v}(\t) = \sup_{y \leq 0}\left\{\t\cdot y - f(y)\right\}.
\]
Since $\bar{v}$ is non-increasing there exists a random variable
$Y(\t) \in L^2(\Prob)$ such that
$-\textnormal{Var}[Y(\t)]\in\partial \bar{v}(\t)$ and the
definition of the subgradient yields
\[
    \bar{v}(\t) = \sup_{Y \in L^2} \left\{\t(-\textnormal{Var}[Y]) -f(-\textnormal{Var}[Y])\right\}.
\]
With the pricing scheme on $L^2(\Prob)$ defined by
\[
    \bar{\pi}(Y) := -\E[Y]-f(-\textnormal{Var}[Y])
\]
this yields
\[
    \bar{v}(\t)=\sup_{Y \in L^2} \left\{U(\t, Y) - \bar{\pi}(Y) \right\}.
\]
The characterization of the subdifferential follows by analogy to part (i).
\end{rmenumerate}
\end{Proof}

The preceding lemma along with Proposition \ref{pr:Uconv} shows that any convex, non-negative and non-increasing function $v$ on $\T$ induces an incentive compatible catalogue $(X,\pi)$ via
\[
    X(\theta) \in \partial_U v(\t) \quad \mbox{and} \quad \pi(\t) = U(\t,X(\t)) - v(\t).
\]
Here we may with no loss of generality assume that $\E[X(\t)] = 0$. In terms of the principal's choice of $v$ her income is given by
\[
   I(v) = \int_{\T} \left( \t v'(\t) - v(\t) \right) d \t.
\]
Since $v \geq 0$ is decreasing and non-negative the principal will
only consider functions that satisfy the normalization constraint
\[
    v(1) = 0.
\]
We denote the class of all convex, non-increasing and non-negative
real-valued functions on $\T$ that satisfy the preceding condition
by ${\cal C}$:
\[
    {\cal C} = \{v: \T \to \re \, \mid \, v \mbox{ is convex, non-increasing, non-negative and } v(1)=0.\}
\]

Conversely, we can associate with any {\bf IC} and ${\bf IR}$
catalogue $(X,\pi)$ a non-negative $U$-convex function of the form
(\ref{v-def2}) where the contract satisfies the variance
constraint $-\textnormal{Var}[X(\t)] = v'(\t)$. In view of the
preceding lemma this function is convex and non-increasing so
after normalization we may assume that $v$ belongs to the class
${\cal C}$. We therefore have the following alternative
formulation of the principal's problem.

\begin{thm}
The principal's optimization problem allows the following alternative formulation:
\begin{equation*}
    \inf \left\{ \varrho \left(W - \int_{\T} X(\t)d\t \right) - I(v) \mid
     v \in {\cal C}, \, \E[X(\theta)] = 0, \, -\textnormal{Var}[X(\t)] = v'(\t) \right\}.
\end{equation*}
\end{thm}

In terms of our alternative formulation we can now prove a
preliminary result. It states that a principal with no initial
endowment will not issue any contracts.

\begin{lemma} If the principal has no initial endowment, i.e., if $W = 0$, then $(v, X)=(0, 0)$ solves her optimization
problem.
\end{lemma}
\begin{Proof}
Since $\varrho$ is a coherent, law invariant risk measure on
$L^2(\Prob)$ that has the Fatou property it satisfies
\begin{equation} \label{lower-bound}
    \varrho(Y) \geq -\E[Y] \quad \mbox{for all } Y \in L^2(\Prob).
\end{equation}
For a given function $v \in {\cal C}$ the normalization constraint
$\E[X(\t)] = 0$ therefore yields
\[
    \varrho\left( - \int_{\T} X(\t) d\t \right) - I(v) \ge
    \E\left[\int_{\T}X(\t)d\t\right] - I(v) = -I(v).
\]
Since $v$ is non-negative and non-increasing $-I(v) \geq 0$.
Taking the infimum in the preceding inequality shows that $v
\equiv 0$ and hence $X(\theta) \equiv 0$ is an optimal solution.
\end{Proof}


\subsection{Minimizing the risk for a given function $v$}

In the general case we approach the principal's problem in two
steps. We start by fixing a function $v$ from the class ${\cal C}$
and minimize the associated risk
\begin{equation*}
    \varrho\left( W - \int_{\T} X(\t)d\t\right)
\end{equation*}
subject to the moment conditions $\E[X(\t)] = 0$ and
$-\textnormal{Var}[X(\t)]=v'(\t)$. To this end, we shall first
prove the existence of optimal contracts $X_v$ for a relaxed
optimization where the variance constraint is replaced by the
weaker condition
\[
    \textnormal{Var}[X(\t)] \leq -v'(\t).
\]
In a subsequent step we show that based on $X_v$ the principal can
transfer risk exposures among the agents in such a way that (i)
the aggregate risk remains unaltered; (ii) the variance constraint
becomes binding. We assume  with no loss of generality that $v$ does not have a jump at $\t=a.$


\subsubsection{The relaxed optimization problem}

For a given $v\in {\cal C}$ let us consider the convex set of derivative securities
\begin{equation}\label{eq:3}
    \X_v:=\left\{X\in\X\,\mid\, E[X(\t)]=0,\,\textnormal{Var}[X(\t)]\le
    -v'(\t)\,\,\mu-a.e.\right\}.
\end{equation}

\begin{lemma} \label{Lemma-bound}
\begin{rmenumerate}
    \item All functions $v \in {\cal C}$ that are acceptable for the principal are uniformly bounded.
    \item Under the conditions of (i) the set $\X_v$ is closed and bounded in $L^2(\Prob \otimes \mu)$. More precisely,
    \[
        \|X\|_2^2\le v(a) \quad{\mbox{for all}}\quad X\in\X_v.
    \]
\end{rmenumerate}
\end{lemma}
\begin{Proof}
\begin{rmenumerate}
    \item If $v$ is acceptable for the principal, then any $X \in \X_v$ satisfies
    \[
        \varrho\left(W-\int_{\T} X(\t)d\t\right) - I(v) \leq \varrho(W).
    \]
    From (\ref{lower-bound}) and that fact that $\E[X(\t)] = 0$ we deduce that
    \[
        - \E[W] - I(v) \leq \varrho\left(W-\int_{\T} X(\t)d\t\right) - I(v) \leq \varrho(W)
    \]
    so
    \[
        -I(v) \leq  \E[W] + \varrho(W) =: K.
    \]
    Integrating by parts twice and using that $v$ is non-increasing and $v(1) = 0$ we see that
    \[
        K \geq -I(v) = a v(a) + 2 \int_a^1 v(\t) d\t \geq a v(a).
    \]
    This proves the assertion because $a > 0$.
    \item For $X \in \X_v$ we deduce from the normalization constraint $v(1) = 0$ that
    \[
        \|X\|^2_2 = \int \int X^2(\t,\omega) d\Prob \, d\t \leq -\int v'(\t) d\t  \leq v(a)
    \]
    so the assertion follows from part (i).
\end{rmenumerate}
\end{Proof}

Since $\varrho$ is a convex risk measure on $L^2$ and because the
set $X_v$ of contingent claims is convex, closed and bounded in
$L^2$ a general result from the theory of convex optimization
yields the following proposition.

\begin{prop}
If the function $v$ is acceptable for the principal, then there
exists a contract $\{X_v(\t)\}$ such that \[
    \inf_{X \in \X_v} \varrho\left(W-\int_{\T} X(\t)d\t\right) = \varrho\left( W - \int_{\T} X_v(\t)d\t\right).
\]
\end{prop}

The contract $X_v$ along with the pricing scheme associated with
$v$ does not yield an incentive compatible catalogue unless the
variance constraints happen to be binding. However, as we are now
going to show, based on $X_v$ the principal can find a
redistribution of risk among the agents such that the resulting
contract satisfies our {\bf IC} condition.

\subsubsection{Redistributing risk exposures among agents}

Let
\begin{equation*}\label{ex:8}
    \partial\X_v=\left\{X\in\X_v\,\mid\, E[X(\t)]=0,\,\textnormal{Var}[X(\t)]=-v'(\t),\,\mu-a.e.\right\}
\end{equation*}
be the set of all contracts from the class $\X_v$ where the
variance constraint is binding. Clearly,
\begin{equation*}\label{ex:9}
    \varrho\left( W - \int_{\T} X_v(\t)d\t\right) \le \inf_{X \in \partial\X_v}
    \varrho\left( W - \int_{\T} X(\t)d\t\right).
\end{equation*}
Let us then introduce the set of types
\begin{equation*}\label{ex:10}
    \T_v:=\left\{\t\in\T\,\mid\, \textnormal{Var}[X_v(\t)]<-v'(\t)\right\},
\end{equation*}
for whom the variance constraint is not binding. If $\mu(\T_v)=0,$
then $X_v$ yields an incentive compatible contract. Otherwise, we
consider a random variable $\tilde{Y}\in\X_v,$ fix some type
$\overline{\t}\in\T$ and define
\begin{equation}\label{ex:11}
Y:=\frac{\tilde{Y}(\overline{\t})}{\sqrt{\textnormal{Var}[\tilde{Y}(\overline{\t})]}}.
\end{equation}

We may with no loss of generality assume that $Y$ is well defined for otherwise the status quo is optimal for the principal and her risk minimization problem is void.
The purpose of introducing $Y$ is to offer a set of structured
products $Z_v$ based on $X_v,$ such that $Z_v$ together with the
pricing scheme associated with $v$ yields an incentive compatible
catalogue. To this end, we choose constants $\tilde{\alpha}(\t)$
for $\t \in \T_v$ such that
\begin{equation*}\label{ex:12}
\textnormal{Var}[X_v(\t)+\tilde{\alpha}(\t)Y]=-v'(\t).
\end{equation*}
This equation holds for
\begin{equation*}\label{ex:13}
\tilde{\alpha}_{\pm}(\t) = -\textnormal{Cov}[X_v(\t), Y] \pm
\sqrt{\textnormal{Cov}^2[X_v(\t),Y]-v'(\t)-\textnormal{Var}[X_v(\t)]}.
\end{equation*}

For a type $\t \in \T_v$ the variance constraint is not binding.
Hence $- v'(\t) -\textnormal{Var}[X_v(\t)] > 0$ so that
$\alpha_+(\t)
> 0$ and $\alpha_-(\t) < 0$. An application of Jensen's
inequality together with the fact that $\|X_v\|_2$ is bounded
shows that $\alpha_\pm$ are $\mu$-integrable functions. Thus there
exists a threshold type $\t^*\in \T$ such that
\begin{equation*}\label{ex:15}
    \int_{\T_v\cap(a,\t^*]}{\alpha}^+(\t)d\t+\int_{\T_v\cap(\t^*,1]}{\alpha}^-(\t)d\t=0.
\end{equation*}
In terms of $\t^*$ let us now define a function
\[
  \alpha(\t):= \left\{
  \begin{array}{ll}
    \tilde{\alpha}^+(\t), & \hbox{if}\quad \t\le\t^*\\
    \tilde{\alpha}^-(\t), & \hbox{if}\quad \t>\t^*
  \end{array}
\right.
\]
and a contract
\begin{equation}\label{ex:16}
Z_v:=X_v+\alpha Y\in\partial\X_v.
\end{equation}
Since $\int \alpha d \t = 0$ the aggregate risks associated with $X_v$ and $Z_v$ are equal. As a result, the contract $Z_v$ solves the risk minimization problem
\begin{equation}\label{ex:17}
    \inf_{X\in\partial\X_v} \varrho \left( W + \int_{\T} X(\t) \mu(d\t) \right).
\end{equation}

\begin{rem}
In Section \ref{sec-examples} we shall consider a situation where
the principal restricts itself to a class of contracts for which
the random variable $X_v$ can be expressed in terms of the
function $v$. In general such a representation will not be
possible since $v$ only imposes a restriction on the contracts'
second moments.
\end{rem}


\subsection{Minimizing the overall risk}

In order to finish the proof of our main result it remains to show that the minimization problem
\[
    \inf_{v \in {\cal C}} \left\{ \varrho \left( W - \int_{\T} Z_v(\theta) \mu(d \t)\right) - I(v)
    \right\}
\]
has a solution and the infimum is obtained. To this end, we
consider a minimizing sequence $\{v_n\} \subset {\cal C}$. The
functions in ${\cal C}$ are locally Lipschitz continuous  because
they are convex. In fact they are {\it uniformly} locally
Lipschitz: by Lemma \ref{Lemma-bound} (i) the functions $v \in {\cal C}$ are uniformly bounded and non-increasing so all the elements of $\partial v (\t)$ are uniformly bounded on compact sets of types.
As a result, $\{v_n\}$ is a sequence of uniformly bounded and uniformly
equicontinuous functions when restricted to compact subsets of
$\T$. Thus there exists a function $\bar{v} \in {\cal C}$ such
that, passing to a subsequence if necessary,
\[
    \lim_{n \to \infty} v_{n} = \bar{v} \quad \mbox{uniformly on compact sets.}
\]
A standard $3\epsilon$-argument shows that the convergence properties of the sequence $\{v_{n}\}$ carry over to the derivatives so that
\[
    \lim_{n \to \infty} v'_{n} = \bar{v}' \quad \mbox{almost surely uniformly on compact sets.}
\]
Since $-\t v'_{n}(\t) + v_{n}(\t) \geq 0$ it follows from Fatou's
lemma that $- I(\bar{v}) \leq \liminf_{n \to \infty} -I(v_{n})$ so
\begin{eqnarray*}
    & & \liminf_{n \to \infty}
    \left\{ \varrho \left( W - \int_{\T} Z_{v_{n}}(\theta) \mu(d \t)\right) - I(v_{n}) \right\} \\
    & \geq & \liminf_{n \to \infty} \varrho \left( W - \int_{\T} Z_{v_{n}}(\theta) \mu(d \t)\right) +
    \liminf_{n \to \infty} - I(v_{n}) \\
    & \geq & \liminf_{n \to \infty} \varrho \left( W - \int_{\T} Z_{v_{n}}(\theta) \mu(d \t)\right) -
    I(\bar{v})
\end{eqnarray*}
and it remains to analyze the associated risk process. For this,
we first observe that for $Z_{v_{n}} \in \partial X_{v_{n}}$
Fubini's theorem yields
\begin{equation} \label{boundary}
    \|Z_{v_{n}}\|^2_2 = \int \int Z^2_{v_{n}} d \Prob d\t = - \int v'_{n}(\t) d \t = v_{n}(a).
\end{equation}
Since all the functions in ${\cal C}$ are uniformly bounded, we
see that the contracts $Z_{v_n}$ are contained in an $L^2$
bounded, convex set. Hence there exists a square integrable random
variable $Z$ such that, after passing to a subsequence if
necessary,
\begin{equation} \label{convergence-Z}
    w-\lim_{n \rightarrow \infty} Z_n = Z 
\end{equation}

Let $Z_{\bar{v}} \in X_{\bar{v}}$. Convergence of the functions $v_n$ implies $\|Z_{v_n}\|_2 \to
\|Z_{\bar{v}}\|_2$. Thus (\ref{convergence-Z}) yields $\|Z\|_2 = \|Z_{\bar{v}}\|_2$ along with convergence of aggregate risks:
\begin{equation*}\label{eq:norm1}
    \|Z\|_2 = \|Z_{\bar{v}}\|_2 \quad \mbox{and} \quad \int_{\T}
    Z_n(\t,\omega) d\t \rightarrow
    \int_{\T} Z(\t,\omega) d\t \quad \mbox{weakly in } L^2(\Prob).
\end{equation*}
By Corollary I.2.2 in Ekeland and T\'{e}mam (1976) \cite{kn:et}, a lower semi-continuous convex function $f:X\to\re$ remains ,so with
respect to the weak topology $\sigma(X, X^*),$ the Fatou
property of the risk measure $\varrho$ guarantees that
\begin{eqnarray*}
    \varrho \left( W - \int_{\T} Z_{\bar{v}}(\theta) \mu(d \t)\right) &\leq&
    \varrho \left( W - \int_{\T} Z(\theta) \mu(d \t)\right) \\
    & \leq & \liminf_{n \to \infty} \varrho \left( W - \int_{\T} Z(\theta) \mu(d
    \t)\right).
\end{eqnarray*}
%
%
%
%
%
%
%
%
%
%
%
We conclude that $(Z_{\bar{v}}, \bar{v})$ solves the Principal's problem.


\section{Examples} \label{sec-examples}

Our main theorem states that the principal's risk minimization problem has a solution. The solution can
be characterized in terms of a convex function that specifies the
agents' net utility. Our existence result is based on a min-max optimization
scheme whose complexity renders a rather
involved numerical analysis . In this section we consider some examples where the
principal's choice of contracts is restricted to class of
numerically more amenable securities. The first example studies a situation
where the principal offers
type-dependent multiplies of some benchmark claim. In this case
the principal's problem can be reduced to a constraint variational
problem that can be solved in closed form. A second example
comprises put options with type dependent strikes. Here we provide
a numerical algorithm for approximating the optimal solution.

\subsection{A single benchmark claim}

In this section we study a model where the principal sells a
type-dependent multiple of a benchmark claim $f(W) \ge 0$ to the
agents. More precisely, the principal offers contracts of
the form
\begin{equation}\label{bond}
    X(\t) = \alpha(\t)f(W).
\end{equation}
In order to simplify the notation we shall assume that the
T-bond's variance is normalized:
\[
    \textnormal{Var}[f(W)]=1.
\]

\subsubsection{The optimization problems}

Let $(X,\pi)$ be a catalogue where the contract $X$ is of the
form (\ref{bond}). By analogy to the general case it will be
convenient to view the agents' optimization problem as an
optimization problem of the set of claims $\{\gamma f(W)
\mid \, \gamma \in \re\}$ so the function $\alpha: \T \rightarrow
\re$ solves
\[
    \sup_{\gamma \in \re} \left\{ U(\t,\gamma f(W)) - \pi(\t) \right\}.
\]
In view of the variance constraint on the agents' claims the
principal's problem can be written as
\begin{equation*}\label{eq:TC1}
    \inf \left\{ \varrho \left( W - C(v)f(W)\right)
    -I(v) \mid v \in {\cal C} \right\} \quad \mbox{where} \quad
    C(v) = \int_{\T}\sqrt{-v'(\t)}d\t.
\end{equation*}

Note that $E[f(W)]>0,$ so the term  $E[f(W)]\sqrt{-v'(\t)}$ must
be included in the income. Before proceeding with the general case
let us first consider a situation where in addition to being
coherent and law invariant, the risk measure $\varrho$ is also
comonotone. In this case each security the principal sells to some
agent increases her risk by the amount
\[
    \varrho\left(-(f(W)-E[f(W)])\sqrt{-v'(\t)}\right)+\left(v(\t)-\t
    v'(\t)\right)\ge 0.
\]
This suggests that it is optimal for the principal not to sell a
bond whose payoff moves into the same direction as her initial
risk exposure.

\begin{prop}\label{pr:comonotone}
Suppose that $\varrho$ is comonotone additive. If $f(W)$ and $W$
are comonotone, then $v=0$ is a solution to the principal's
problem.
\end{prop}
\begin{Proof}
If $W$ and $f(W)$ are comonotone,  then the risk measure in
equation (\ref{eq:TC1}) is additive and the principal needs to
solve
\begin{equation*}\label{eq:TC}
    \varrho \left( W \right) + \inf_{v \in {\cal C}}
    \int_{\T}\left(v(\t)+\varrho\left(f(W)-\E[f(W)]\right)\sqrt{-v'(\t)}-\t
    v'(\t)\right)d\t.
\end{equation*}
Since $\varrho\left(f(W)-E[f(W)]\right)\ge 0$ and $-\t v'(\t)\ge
0$ we see that
\[
    \int_{\T}\left(v(\t)+\rho\left(F(W)\right)\sqrt{-v'(\t)}-\t
    v'(\t)\right)d\t\ge 0
\]
and hence $v\equiv 0$ is a minimizer.
\end{Proof}

In view of the preceding proposition the principal needs to design
the payoff function $f$ in such a way that $W$ and $f(W)$ are not
comonotone. We construct an optimal payoff function in the
following subsection.

\subsubsection{A solution to the principal's problem}

Considering the fact that $\varrho(\cdot)$ is a decreasing
function the principal's goal must be to make the quantity $C(v)$
as small as possible while keeping the income as large as
possible. In a first step we therefore solve, for any constant $A
\in \re$ the optimization problem
\begin{equation} \label{variational-problem}
    \sup_{v \in {\cal C}} \, C(v) \quad \mbox{subject to} \quad
    \int_{\T}\left(\E[f(W)] \sqrt{-v'(\t)}-v(\t)+\t
    v'(\t)\right)d\t=A.
\end{equation}

The constraint variational problem (\ref{variational-problem}) captures the problem of risk minimizing subject to an income constraint. It
can be solved in closed form. The associated Euler-Lagrange equation
is given by
\begin{equation}\label{eq:TC4}
 \lambda=\frac{d}{d\t}\left(-\lambda \t+\frac{\lambda \E[f]-1}{2\sqrt{-v'(\t)}}\right),
\end{equation}
where $\lambda$ is the Lagrange multiplier. The income constraint
and boundary conditions are:
\[
    v'(a)=-\frac{(\lambda')^2}{4\lambda^2 a^2} \quad\mbox{and}\quad
    v(1)=0 \quad \mbox{where} \quad \lambda' = (\lambda \E[f]-1).
\]

Integrating both sides of equation (\ref{eq:TC4}) and taking into
account the normalization condition $v(1)=0$, we obtain
\begin{equation*}\label{eq:TC5}
    v(\t)=\frac{1}{8}\left(\frac{\lambda'}{\lambda}\right)^2\left[\frac{1}{2\t-a}-\frac{1}{2
    -a}\right].
\end{equation*}
Inserting this equation into the constraint yields
\[
    A=\E[f]\sqrt{\left(\frac{\lambda'}{\lambda}\right)^2}\int_a^1
    \frac{d\t}{2\t-a}-\left(\frac{\lambda'}{\lambda}\right)^2\int_a^1\left\{\frac{1}{8}\left[\frac{1}{2\t-a}-\frac{1}{2
    -a}\right]+ \frac{1}{4}\frac{\t}{(2\t-a)^2}\right\}d\t.
\]
In terms of
\[
    M:=\int_a^1\left\{\frac{1}{8}\left[\frac{1}{2\t-a}-\frac{1}{2
    -a}\right]+ \frac{1}{4}\frac{\t}{(2\t-a)^2}\right\}d\t \quad
    \mbox{and} \quad N:=\int_a^1 \frac{d\t}{2\t-a}
\]
we have the quadratic equation
\[
    -M\left(\frac{\lambda'}{\lambda}\right)^2+N \E[f] \sqrt{\left(\frac{\lambda'}{\lambda}\right)^2}-A=0,
\]
which has the solution
\begin{equation*}\label{eq:toy5}
\sqrt{\left(\frac{\lambda'}{\lambda}\right)^2}=\frac{N
\E[f]-\sqrt{(N \E[f])^2-4AM}}{2M}
\end{equation*}
We have used the root with alternating signs, as we require the
problem to reduce to $\varrho(W)$ for $A=0.$

\begin{rem}
We notice that the constraint variational problem
(\ref{variational-problem}) is independent of the risk measure
employed by the principal. This is because we minimized the risk
pointwise subject to a constraint on aggregate revenues.
\end{rem}

In view of the preceding considerations the principal's problem
reduces to a one-dimensional minimization problems over the Reals:
\[
    \inf_A
    \varrho\left( W - f(W)\frac{N^2 \E[f]}{2M}+
    f(W)\frac{N}{2M}\sqrt{(N \E[f])^2 -4A M}\right)-A.
\]
Once the optimal value $A^*$ has been determined, the principal
offers the securities
\[
    \left(\frac{\lambda \E[f] -1}{4\t\lambda-2\lambda
    a}\right)f(W)
\]
at a price
\[
    \frac{\lambda \E[f] -1}{2}\left(\frac{3E\lambda
    (2\t-a)-a}{(4\t\lambda-2\lambda a)^2}+\frac{\lambda E
    -1}{\lambda^2}\frac{1}{2-a}\right).
\]

\begin{example}
Assume that the principal measures her risk exposure using Average
Value at Risk at level $0.05$. Let $\tilde{W}$ be a normally
distributed random variable with mean $1/2$ and variance $1/20.$ One can think that $\tilde{W}$ represents temperature.
Suppose that the principal's initial income is exposed to temperature risk and it  is given by $W =
0.1(\tilde{W}-1.1)$ with associated risk
\[
    \varrho(W)=0.0612.
\]
Suppose furthermore that the principal sells units of a put option
on $\tilde{W}$ with strike $0.5$, i.e.,
\[
    f(W)=(W-0.5)^+
\]
By proceeding as above we approximated the principal's risk as
$-0.6731$ and she offers the security
\[
    X(\t)=\frac{0.5459}{2\t-a}f(W)
\]
to the agent of type $\t$ for a price
\[
    \pi(\t)=\frac{1.1921}{8(2-a)}-\frac{(1.1921)\t-(0.22)(2\t-a)}{\sqrt{2}(2\t-a)^2}.
\]
\end{example}


\subsection{Put options with type dependent strikes}

In this section we consider the case where the principal
underwrites put options on her income with type-dependent strikes.
We assume that $W \leq 0$ is a bounded random variable and
consider contracts of the form
\[
    X(\t) = (K(\t) - |W|)^+ \quad \mbox{with} \quad 0 \leq K(\t)
    \leq \|W\|_\infty.
\]

The boundedness assumption on the strikes is made with no loss of
generality and each equilibrium pricing scheme is necessarily
non-negative. Note that in this case the risk measure can be
defined on $\ml(\Prob),$ so we only require convergence in
probability to use the Fatou property. We deduce that both the
agents' net utilities and the variance of their positions is
bounded from above by some constants $K_1$ and $K_2$,
respectively. Thus, the principal chooses a function $v$ and
contract $X$ from the set
\[
    \{(X,v) \mid v \in {\cal C}, \, v \leq K_1, \, -\textnormal{Var}[K(\t)-|W|] =
    v'(\t), \, |v'| \leq K_2, \, 0 \leq K(\t) \leq \|W\|_\infty \}.
\]

The variance constraint $v'(\t)= -\textnormal{Var} [(K(\t)-W)^+]$
allows us to express the strikes in terms of a continuous function
of $v'$, i.e.,
\[
    K(\t) = F(v'(\t)).
\]
The Principal's problem can therefore be written as
\[
    \inf \left\{
    \varrho\left( W - \int_{\T}\left\{ (F(v'(\t))-|W|)^+ - \E[(F(v'(\t))-|W|)^+] \right\} d\right) -
    I(v) \right\}
\]
where the infimum is taken over the set of all functions $v \in
{\cal C}$ that satisfy $v \leq K_1$ and $|v'| \leq K_2$.

\begin{rem}
Within our current framework the contracts are expressed in terms
of the derivative of the principal's choice of $v$. This reflects
the fact that the principal restricts itself to type-dependent put
options and is not always true in the general case.
\end{rem}


\subsubsection{An existence result}

Let $\{v_n\}$  be a minimizing sequence for the principal's
optimization problem. The functions $v_n$ are uniformly bounded
and uniformly equicontinuous so we may with no loss of generality
assume that $v_n \to \overline{v}$ uniformly. Recall this also implies
a.s. convergence of the derivatives. By dominated
convergence and the continuity of $F,$ along with the fact that $W$ is
bounded yields
\begin{equation*}\label{eq:fatou}
    \int_{\T}(F(v_n'(\t))-|W|)^+d\longrightarrow\int_{\T}(F(\overline{v}'(\t))-|W|)^+d
    \quad \Prob{\mbox{-a.s}}.
\end{equation*}
and
\begin{equation*}\label{eq:DCT}
    \lim_{n\to\infty} \int_{\T} \E[(F(v_n'(\t))-|W|)^+d\t =
    \int_{\T} \E[(F(\overline{v}'(\t))-|W|)^+d\t
\end{equation*}
This shows that the principal's positions converge almost surely
and hence in probability. Since $\varrho$ is lower-semi-continuous
with respect to convergence in probability we deduce that
$\overline{v}$ solves the principal's problem.


\subsubsection{An algorithm for approximating the optimal solution}{\label{NumEx}}

We close this paper with a numerical approximation scheme for the principal's optimal solution within the pit option framework. We assume the set of states of the World is finite
with cardinality $m.$  Each possible state $\omega_j$  can occurs
with probability $p_j.$ The realizations of the principal's wealth
are denoted by $W=(W_1,\ldots,W_m).$ Note that $p$ and $W$ are
treated as known data. We implement a numerical algorithm to
approximate a solution to the principal's problem when she
evaluates risk via the risk measure
\begin{equation*}
\varrho(X)=-\sup_{q\in Q_{\lambda}}\sum_{j=1}^m X(\omega_j)p_j q_j,
\end{equation*}
where
\begin{equation*} Q_{\lambda}:=\left\{q\in\re^m_+\,\mid\,
p\cdot q=1,\,q_j\le\lambda^{-1}\right\}.
\end{equation*}

We also assume the set of agent types is finite with cardinality
$n,$ i.e. $\t=(\t_1,\ldots,\t_n).$ The density of the types is
given by $M:=(M_1,\ldots, M_n).$ In order to avoid singular points
in the principal's objective function, we approximate the option's
payoff function $f(x)=(K-x)^+$ by the differentiable function
\[
\begin{array}{cc}
  T (x, K)= & \left\{
  \begin{array}{lll}
    0, & \hbox{if}\quad x\le K-\epsilon, \\
    S(x,K), & \hbox{if}\quad K-\epsilon<x<K+\epsilon,\\
    x-K, & \hbox{if}\quad x\ge K+\epsilon.
  \end{array}
\right.
\end{array}
\]
where
\[
    S(x,K)=\frac{x^2}{4\epsilon}+\frac{\epsilon-K}{2\epsilon}x+\frac{K^2-2A\epsilon+\epsilon^2}{4\epsilon}.
\]

The algorithm uses a penalized Quasi-Newton method, based on  Zakovic and  Pantelides \cite{kn:zp}, to approximate a
minimax point of
\begin{eqnarray*}
    F(v,K,q) &=& -\sum_{i=1}^nW_ip_iq_i+\frac{1}{n}\sum_{i=1}^n\left(\sum_{j=1}^n
    T(K_j-|W_i|)\right)p_iq_i-\frac{1}{n}\sum_{i=1}^n\left(\sum_{j=1}^{n-1}
    T(K_j-|W_i|)\right)p_i\\
    & & +\frac{1}{n}\sum_{i=1}^n\left(
    v_i-\t_i\frac{v_{i+1}-v_i}{\t_{i+1}-\t_i}\right)+\frac{1}{n}\left(v_n-\frac{v_n-v_{n-1}}{1-\t_{n-1}}\right)
\end{eqnarray*}
where  $v=(v_1,\ldots,v_n)$ stands for the values of a convex,
non-increasing function, $K=(K_1,\ldots,K_n)$ denotes the vector
of type dependent strikes and the derivatives $v'(\t_i)$ are
approximated by
\[
    v'(\t_i) = \frac{v_{i+1}-v_i}{\t_{i+1}-\t_i}.
\]

The need for a penalty method arises from the fact that we face
the equality constraints $v'(\t)=-Var[(K(\t)-|W|)^+]$ and $p\cdot
q=1.$ In order to implement a descent method, these constraints
are relaxed and a penalty term is added. We denote by $ng$ the
total number of constraints.  The principal's problem is to find
\[
    \min_{(v,K)}\max_{q\in Q_{\lambda}} F(v, K, q) \quad \mbox{subject
    to} \quad G(v,K,q)\le 0
\]
where $G:\re^{2n+m}\to\re^{ng}$ determines the constraints that
keep $(v,K)$ within the set of feasible contracts and $q\in
Q_{\lambda}.$ The Maple code for our procedure is given in the appendix for completeness.

\begin{example}
Let us illustrate the effects of risk transfer on the principal's position in two model with five agent types and two states of the world. In both cases $W=(-1, -2),$  $\t=(1/2, 5/8, 3/4, 7/8, 1)$ and $\lambda=1.1$ The starting values  $v_0,$ $q_0$ and $K_0$ we set are $(4,3,2,1,0),$ $(1,1)$ and $(1,1,1,1,1)$ respectively.

\begin{itemize}
\item[i)] Let $p=(0.5, 0.5)$ and the types be uniformly distributed. The principal's
initial evaluation of her risk is $1.52$. The optimal function $v$ and strikes are:

\begin{center}
\begin{tabular}{|c|c|}
   \hline
  $V_{1}$ & 0.1055 \\\hline
  $V_{2}$ & 0.0761 \\\hline
  $V_{3}$ & 0.0501 \\\hline
  $V_{4}$ & 0.0342 \\\hline
  $V_{5}$ & 0.0195 \\ \hline

\end{tabular}\hspace{1in}
\begin{tabular}{|c|c|}
   \hline
  $K_{1}$ & 1.44 \\\hline
  $K_{2}$ & 1.37 \\\hline
  $K_{3}$ & 1.07 \\\hline
  $K_{4}$ & 1.05 \\\hline
  $K_{5}$ & 1.05 \\ \hline
\end{tabular}\end{center}

\noindent The Principal's valuation of her risk after the exchanges
with the agents decreases to  $0.2279.$

\begin{figure}[ht!]
\begin{center}
\subfigure[\label{figure:strike1} The type-dependent strikes.]
{\epsfig{figure=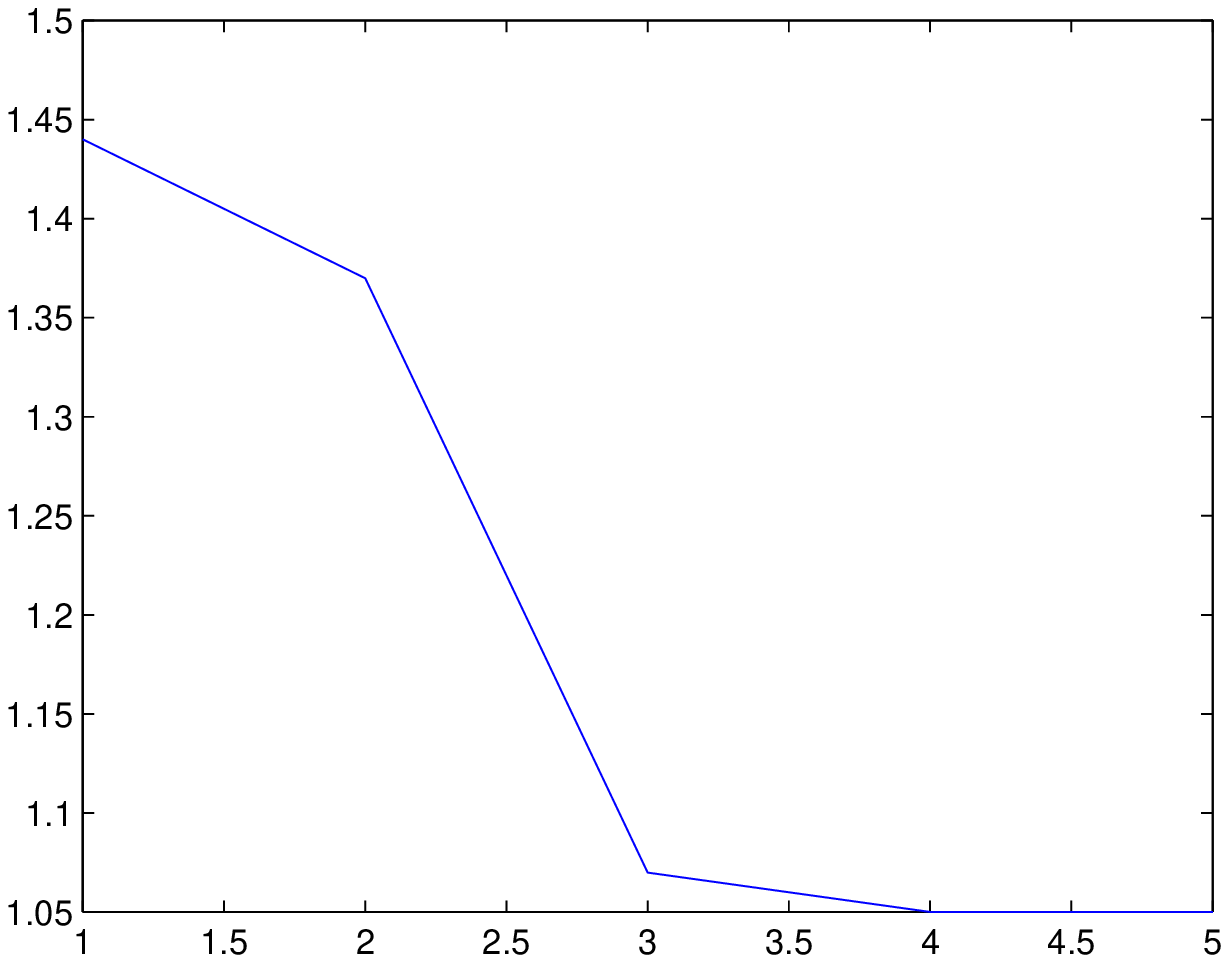,scale=0.45}} \hspace{5mm}
\subfigure[\label{figure:v1}The optimal function $v$.]
{\epsfig{figure=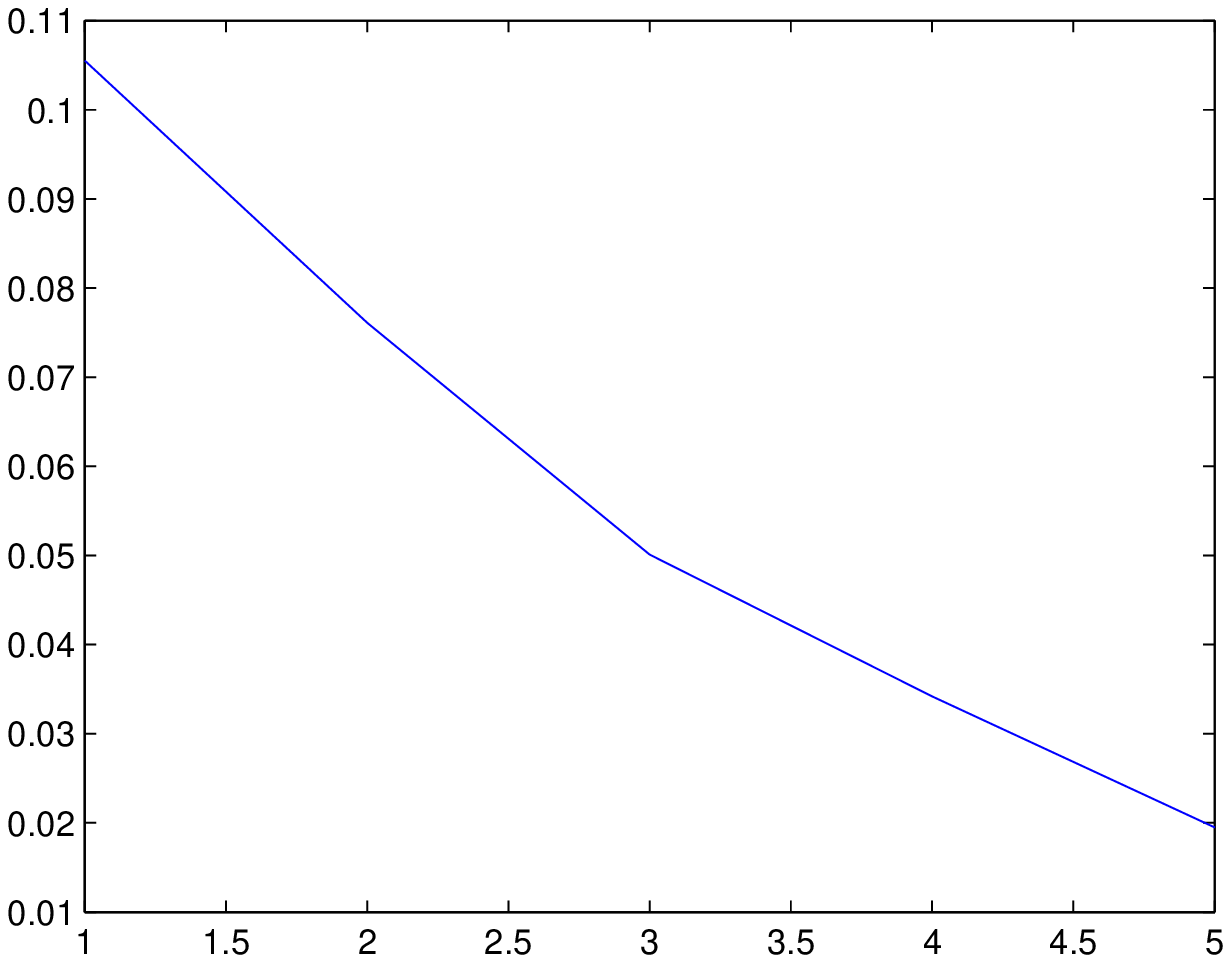,scale=0.45}} \caption{\label{figure:1}
Optimal solution for underwriting put options, Case 1.}
\end{center}
\end{figure}

\item[ii)] In this instance $p=(0.25, 0.75)$ and $M=(1/15, 2/15, 3/15, 4/15, 5/15).$
The principal's initial evaluation of her risk is $1.825$. The values for the discretized  $v$ the type-dependent strikes are:

\begin{center}
\begin{tabular}{|c|c|}
   \hline
  $V_{1}$ &  0.0073\\\hline
  $V_{2}$ &  0.0045\\\hline
  $V_{3}$ &  0.0029\\\hline
  $V_{4}$ &  0.0026\\\hline
  $V_{5}$ &  0.0025\\ \hline

\end{tabular}\hspace{1in}
\begin{tabular}{|c|c|}
   \hline
  $K_{1}$ &  1.27\\\hline
  $K_{2}$ &  1.16\\\hline
  $K_{3}$ &  1.34\\\hline
  $K_{4}$ &  0.11\\\hline
  $K_{5}$ &  0.12\\ \hline
\end{tabular}
\end{center}

\noindent The Principal's valuation of her risk after the exchanges
with the agents is $0.0922.$

\begin{figure}[ht!]
\begin{center}
\subfigure[\label{figure:strike2} The type-dependent strikes.]
{\epsfig{figure=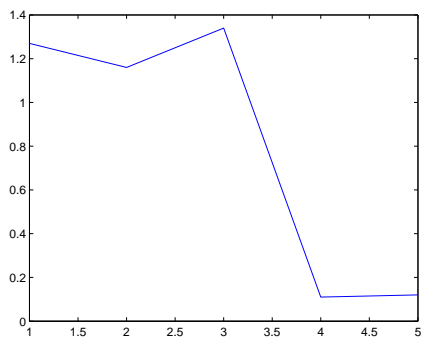,scale=0.45}} \hspace{5mm}
\subfigure[\label{figure:v2}The optimal function $v$.]
{\epsfig{figure=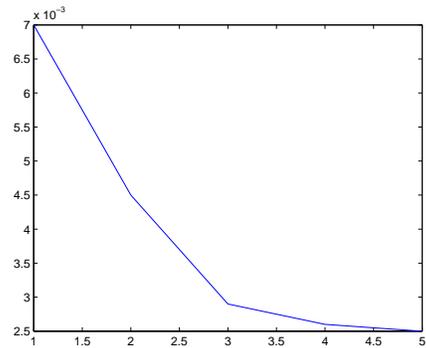,scale=0.45}} \caption{\label{figure:2}
Optimal solution for underwriting put options, Case 2.}
\end{center}
\end{figure}

\end{itemize}

\end{example}

\section{Conclusions}

In this paper we analyzed a screening problem where the principal's choice space is infinite dimensional. Our motivation was to present a nonlinear pricing scheme for over-the-counter financial products, which she trades with a set of heterogeneous agents with the aim of minimizing the exposure of her income to some non-hedgeable risk. In order to characterize incentive compatible and individually rational catalogues, we have made use of U-convex analysis. To keep the problem tracktable we have assumed the agents have mean-variance utilities, but this is not necessary for the characterization of the problem. Considering more general utility functions is an obvious extension to this work. Our main result is a proof of existence of a solution to the principal's risk minimization problem in a general setting. The examples we have studied suggest that the methodologies for approaching particular cases are highly dependent on the choice of risk measure, as well as on the kinds of contracts the principal is willing (or able) to offer.  In most cases obtaining closed form solutions is not possible and implementations must be done using numerical methods. As a work in progress we are considering agents with heterogenous initial endowments (or risk exposures), as well as a model that contemplates an economy with multiple principals.  

\begin{appendix}

\section{Coherent risk measures on $L^2$.}

In this appendix we recall some properties and representation
results for risk measures on $L^2$ spaces; we refer to the textbook of F\"ollmer and Schied \cite{kn:fs} for a detailed discussion of convex risk measures on $L^\infty$ and to Cheridito and Tianbui \cite{kn:cht} for risk measures on rather general state spaces. B\"auerle and M\"uller \cite{kn:ba} establish representation properties of risk law invariant risk measures on $L^p$ spaces for $p \geq 1$. We assume that all
random variables are defined on some standard non-atomic
probability space $(\Omega,\mf,\Prob)$.

\begin{definition}
\begin{rmenumerate}
\item A {\it{monetary measure of risk}} on $L^2$ is a function
$\varrho:L^{2}\to\re\cup \{\infty\}$ such that for all $X,Y \in
L^{2}$ the following conditions are satisfied:
\begin{itemize}
\item Monotonicity: if $X\le Y$ then $\varrho(X)\ge\varrho(Y)$.

\item Cash Invariance: if $m\in\re$ then $\varrho(X+m)=\varrho(X)-m$.
\end{itemize}
\item A risk measure is called {\it coherent} if it is convex and homogeneous of degree 1, i.e., if the following two conditions hold:
\begin{itemize}
\item Convexity: for all $\lambda \in [0,1]$ and all positions
$X,Y \in L^2$:
\[
    \varrho(\lambda X+(1-\lambda)Y)\le\lambda\varrho(X)+(1-\lambda)\varrho(Y)
\]

\item Positive Homogeneity: For all $\lambda \geq 1$
\[
    \varrho(\lambda X)=\lambda\varrho(X).
\]
\end{itemize}
\item The risk measure is called coherent and law invariant, if, in addition,
\[
    \rho(X)=\rho(Y)
\]
for any two random variables $X$ and $Y$ which have the same law.
\item The risk measure $\varrho$ on $L^2$ has the {\it Fatou
property} if for any sequence of random variables $X_1,
X_2,\ldots$ that converges in $L^2$ to a random variable $X$ we
have
\[
    \rho(X)\le\liminf_{n\to\infty} \rho(X_n).
\]
\end{rmenumerate}
\end{definition}

Given $\lambda\in (0,1],$ the \textsl{Average Value at Risk} of
level $\lambda$ of a position $Y$ is defined as
\[
    AV@R_{\lambda}(Y):=-\frac{1}{\lambda}\int_0^{\lambda} q_Y(t)dt,
\]
where $q_Y(t)$ is the upper quantile function of $Y$. If $Y \in
L^{\infty}$, then we have the following characterization
\[
    AV@R_{\lambda}(Y)=\sup_{Q\in{\cal{Q}}_{\lambda}}-\E_Q[Y]
\]
where
\[
    {\cal{Q}}_{\lambda}=\left\{Q<<P\,\mid\,\frac{dQ}{dP}\le\frac{1}{\lambda}\right\}.
\]

\begin{prop}\label{pr:1} For a given financial position $Y \in L^2$
the mapping $\lambda \mapsto AV@R_{\lambda}(Y)$ is decreasing in $\lambda.$
\end{prop}

It turns out the Average Value of Risk can be viewed as a basis
for the space of all law-invariant, coherent risk measures with
the Fatou property. More precisely, we have the following result.

\begin{thm}\label{th:avar} The risk measure
$\varrho: L^{2} \to \re$ is law-invariant, coherent and has the
Fatou Property if and only if $\varrho$ admits a representation of
the following form:
\[
    \varrho(Y)=\sup_{\mu\in M}\left\{\int_0^1
    AV@R_{\lambda}(Y)\mu(d\lambda)\right\}
\]
where $M$ is a set of probability measures on the unit interval.
\end{thm}

As a consequence of Proposition \ref{pr:1} and Theorem
\ref{th:avar} we have the following Corollary:

\begin{cor}\label{cor:coh}
If $\varrho: L^2\to\re$ is a law-invariant, coherent risk measure
with the Fatou Property then
\[
    \varrho(Y)\ge -\E[Y].
\]
\end{cor}

An important class of risk measures are comonotone risk measures
risk. Comonotone risk measures are characterized by the fact that
the risk associated with two position whose payoff ``moves in the
same direction'' is additive.

\begin{definition}  A risk measure $\varrho$ is said to be  {\it {comonotone}} if
\[
    \varrho(X+Y)=\rho(X)+\rho(Y)
\]
whenever $X$ and $Y$ are comonotone, i.e., whenever
\[
    (X(\omega)-X(\omega'))(Y(\omega)-Y(\omega'))\ge 0 \quad
    \Prob\mbox{-a.s.}
\]
\end{definition}

Comonotone, law invariant and coherent risk measures with the Fatou
property admit a representation of the form
\[
    \varrho(Y) = \int_0^1 AV@R_{\lambda}(Y)\mu(d\lambda).
\]

\section{Maple code for the example of Section \ref{NumEx}}
{\small

\noindent with(LinearAlgebra)\\
\noindent n := 5:\\\noindent m := 2:\\\noindent ng := 2*m+4*n+1:\\

\noindent This section constructs the objective function  $f$ and its gradient.\\

\noindent x := Vector(2*n, symbol = xs):\\\noindent q := Vector(m, symbol = qs):\\

$
\begin{array}{ll}
  f := & add(-G[j]*p[j]*q[j], j = 1 .. m)+\\
   & add(add(T(x[i+n],W[j])*M[i]*p[j]*q[j], i = 1 .. n), j = 1 .. m)- \\
  & add(add(T(x[i+n],W[j])*M[i]*p[j], i = 1 .. n), j = 1 ..m)+ \\
   & add((x[i]-t[i]*(x[i+1]-x[i])/(t[i+1]-t[i]))*M[i], i = 1..n-1)+\\
  & x[n]-t[n]*(x[n]-x[n-1])*M[n]/(t[n]-t[n-1]):\\
\end{array}
$

\noindent$ TT := (x, K)->
1/4*x^2/eps+1/2*(eps-K)*x/eps+1/4*(K^2-2*K*eps+eps^2)/eps:$\\
\noindent $T := (x, K) -> piecewise(x \le K-eps, 0, x <
K+eps, TT(x, K), K+eps\le x, x-K):$\\

\noindent gradfx := 0:\\\noindent gradfq := 0:\\
\noindent for i from 1 to 2*n \\do gradfx[i] := diff(f, x[i]) \\end do:\\
\noindent for i from 1 to m\\ do gradfq[i] := diff(f, q[i])\\ end do:\\

\noindent This section constructs the constraint function  $g$ and  its gradient.\\

\noindent g := Vector(ng, symbol = tt):\\
\noindent for j from 1 to n\\ do g[j] := -x[j] end do:\\
\noindent for j from 1 to n-1\\ do g[j+n] :=
x[j+1]-x[j]\\ end do:\\
\noindent for i from 1 to n-1\\ do g[i+2*n-1] := add(T(x[i+n],
$W[j])^2*p[j], j = 1 .. m)-add(T(x[i+n], W[j])*p[j], j = 1 ..
m)^2+(x[i+1]-x[i])/(t[i+1]-t[i])-eps2$\\ end do:\\
\noindent$g[3*n-1] := add(T(x[2*n], W[j])^2*p[j], j = 1 ..
m)-add(T(x[2*n],
W[j])*p[j], j = 1 .. m)^2+(x[n]-x[n-1])/(t[n]-t[n-1])-eps2:$\\
\noindent for i from 1 to n-1\\
 do $g[i+3*n-1] := -add(T(x[i+n], W[j])^2*p[j],
j = 1 .. m)+add(T(x[i+n], W[j])*p[j], j = 1 ..
m)^2-(x[i+1]-x[i])/(t[i+1]-t[i])-eps2$\\ end do:\\
\noindent$g[4*n-1] := -add(T(x[2*n], W[j])^2*p[j], j = 1 ..
m)+add(T(x[2*n],
W[j])*p[j], j = 1 .. m)^2-(x[n]-x[n-1])/(t[n]-t[n-1])-eps2:$\\
\noindent g[4*n] := add(p[i]*q[i], i = 1 .. m)-1+eps3:\\
\noindent g[4*n+1] := -add(p[i]*q[i], i = 1 .. m)-1-eps3:\\
\noindent for i from 1 to m\\
 do g[i+4*n+1] := -q[i]\\ end do:\\
\noindent for i to m\\ do g[i+m+4*n+1] := q[i]-lambda\\ end do:
\noindent gradgx := 0:\\\noindent gradgq := 0:\\

\noindent  for i from 1 to ng\\ do for j from 1 to 2*n\\
 do gradgx[i, j] := diff(g[i], x[j])\\
 end do:\\
 end do:\\
\noindent for i from 1 to ng\\
do for j from 1 to m\\
 do gradgq[i, j] := diff(g[i], q[j])\\
 end do:\\
end do:\\

\noindent This section  constructs the slackness structures.\\

\noindent e := Vector(ng, 1):\\
s := Vector(ng, symbol = si):\\
 z := Vector(ng, symbol = zi):\\
  S :=DiagonalMatrix(s):\\
  Z := DiagonalMatrix(z):\\

\noindent This section  initializes the variable and parameter vectors.\\

\noindent x := Vector(2*n, symbol = xs):\\
 q := Vector(m, symbol = qs):\\
  p :=Vector(m, symbol = ps):\\
   W := Vector(m, symbol = ws):\\
    G := Vector(m,symbol = gs):\\
     t := Vector(n, symbol = ts):\\
     M := Vector(n, symbol = ms):\\
chi := convert([x, q, s, z], Vector):\\

\noindent This section constructs the Lagrangian and its Hessian
matrix.\\

\noindent F
:=convert([gradfx+(VectorCalculus[DotProduct])(Transpose(gradgx),
z),\\ gradfq-(VectorCalculus[DotProduct])(Transpose(gradgq), z),\\
(VectorCalculus[DotProduct])((VectorCalculus[DotProduct])(Z, S),
e)-mu*e, g+s], Vector):\\

\noindent DF := 0:\\
for i from 1 to 2*n+m+2*ng\\
 do for j from 1 to 2*n+m+2*ng\\
 do DF[i, j] :=diff(F[i], chi[j])\\
 end do:\\
  end do:\\

\noindent This section inputs the initial values of the variables and the values of the parameters.\\

\noindent xinit := (4,3, 2, 1, 0, 1, 1, 1, 1, 1):\\
qinit := (1, 1):\\
 sinit := (.1, .1, .1, .1, .1, .1, .1, .1,
.1, .1, .1, .1, .1, .1, .1, .1, .1, .1, .1, .1, .1, .1, .1, .1,
.1):\\
 zinit := (1, 1, 1, 1, 1, 1, 1, 1, 1, 1, 1, 1, 1, 1, 1,
1, 1, 1, 1, 1, 1, 1, 1, 1, 1):\\
 tinit := (1/2, 5/8, 3/4, 7/8,1):\\
  pinit := (.5, .5):\\
   ginit := (-1, -2):\\
    minit := (1,1, 1, 1, 1):\\
     winit := (1, 2):\\
      tau := .1; mu := .1; rho := .5: lambda := 1.1: eps := .1: eps2 := .1: eps3 := .2:\\

\noindent xo := xinit; qo :=qinit; so := sinit; zo := zinit; po :=
pinit; go := ginit; mo :=
minit; wo := winit;\\

\noindent xs := xinit; qs := qinit; si := sinit; zi := zinit; ps :=
pinit; gs := ginit; ms := minit; ws := winit; ts :=
tinit:\\

\noindent The following section contains the executable code.\\

\noindent i := 0; j := 0\\
 normF:=Norm(F,2):\\
  while  (normF $> mu$ and $j<40$)\\
     do\\
     printf("  Inner loop: \#Iteration = \%g /n",j);\\
     printf("    - Solve Linear System (11) ...");\\
       d := LinearSolve(DF, -Transpose(F)):\\
       printf("done\n");\\
     printf("    - Update Params ...");\\
       alphaS:=min(seq(select(type,-s[k]/d[2*n+m+k],positive), k = 1 ..
       Dimension(s)));\\
       alphaZ:=min(seq(select(type,-z[k]/d[2*n+m+ng+k],positive), k = 1 ..
       Dimension(z)));\\
       alphamax:=min(alphaS,alphaZ):\\
       alphamax:=min(tau*alphamax,1);\\
       printf("done/n");\\
       chiold := chi;\\
       chinew := chiold+alphamax*d;\\
       xs := chinew[1 .. 2*n]: ys := chinew[2*n+1 .. 2*n+m]:\\
       si := chinew[2*n+m+1 .. 2*n+m+ng]: zi := chinew[2*n+m+ng+1 ..
       2*n+m+2*ng]:\\
       normF:=Norm(F,2);\\
       printf("done\n");\\
     j:=j+1;\\
end do:
}
\end{appendix}

\end{document}